# Evaluation of Lifetime Bounds
# of Wireless Sensor Networks

Moslem Amiri

Masaryk University, Faculty of Informatics, Botanicka 68a,
602 00 Brno, Czech Republic
amiri@mail.muni.cz

**Abstract.** In this paper we estimate lifetime bounds of a network of motes which communicate with each other using IEEE 802.15.4 standard. Different frame structures of IEEE 802.15.4 along with CSMA/CA medium access mechanism are investigated to discover the overhead of channel acquisition, header and footer of data frame, and transfer reliability during packet transmission. This overhead makes the fixed component, and the data payload makes the incremental component of a linear equation to estimate the power consumed during every packet transmission. Finally we input this per-packet power consumption in a mathematical model which estimates the lower and upper bounds of routings in the network. We also implemented a series of measurements on CC2420 radio used in a wide range of sensor motes to find the fixed and incremental components, and finally the lifetime of a network composed of the motes using this radio.

## 1 Introduction

A Wireless Sensor Network can be composed of tens to thousands of sensor motes spread in a wide area and communicating with each other either directly or through the other nodes. One or more of the nodes are the base station with more or unlimited power supply than the other nodes. The user either has direct access to the base station or through a wired network. The motes have limited power supply without the possibility to be recharged. Most of their power supply is consumed for their transmission purposes and a small fraction is used up during internal processing or sensing operation. According to [1], one bit transmitted in WSNs consumes about as much power as executing 800-1000 instructions. Thus, power consumption of routings is large enough to overshadow the power consumed by the other operations.

In this paper we focus on the power consumption of routings to decide the lifetime of a WSN. The paper is organized as follows. In section 2, we summarize a mathematical model of energy consumption of routings presented earlier. Next in section 3, per-packet energy consumption of routings based on the IEEE 802.15.4 standard is presented and the lifetime bounds of a network based on this standard is suggested. Then in section 4, we represent some practical power measurements of CC2420 radio during different operations, and rewrite the equations gained in section 3 specifically for CC2420 radio. Later in this section, we explain in detail how to calculate the bounds through an example. Section 5 presents the conclusions.

## 2 A Mathematical Model of Energy Consumption of Routings

In order to measure the lower and upper bounds of the energy consumption of routings, we use a mathematical model suggested by Alonso et al. [2]. This model considers continuous sensor networks [3] in which "the sensors communicate their data continuously at a pre-specified rate". In such a network, sensor nodes read sensor values, send them in a multi-hop fashion to a base station and sleep until the next iteration. The leaf sensors send only their own values, while the inner nodes send their own values along with the packets originating from the outer nodes. Received packets are retransmitted unchanged to the base station i.e. nodes do not aggregate the data payload of a received packet with that of another received packet or with their own data payload. The nodes iterate this process till the end of the lifetime of the network. By the term "lifetime of the network", we mean the duration of the time from the initialization of the network till at least one of the nodes dies.

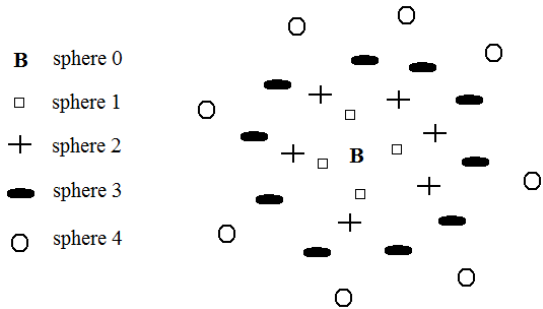

**Fig. 1.** A sensor network partitioned in spheres.

In this model, the set of all nodes ($V$) is partitioned into different subsets (called spheres), $S_0, S_1, \ldots, S_k$ such that $V = S_0 \cup S_1 \cup \ldots \cup S_k$ and $S_i \cap S_j = \emptyset$ for all $i \neq j$ and no $S_i$ is empty. $S_i$ is the set of nodes that can be reached from the base station ($B$) in $i$ hops (thus $S_0 = \{B\}$), but not less than $i$ hops. Figure 1 shows an example of such network. This model assumes that all nodes transmit at the same constant power.

The term *balls* of radius $i$ denoted $B_i$ is introduced in this model such that $B_i = S_0 \cup ... \cup S_i$. Some other definitions are: $s_i = |S_i|, b_i = |B_i|, N = |V|$. Using these definitions, a lower bound on the energy consumption for a node in $S_i$ is:

$$m_i = \frac{N - b_i}{s_i} r + \frac{N - b_i + s_i}{s_i} t \qquad (1)$$

where $r$ is the energy consumed for receiving one packet and $t$ is the energy consumed for transmitting one packet. In the equation above, $N - b_i$ is the total number of nodes outside $B_i$ and thus the total number of packets that the set of nodes in sphere $S_i$ receive in each iteration. The nodes in $S_i$ must transmit all the packets they receive plus their own packets. The minimum energy consumption is when the total number of packets received and transmitted is equally divided among the nodes in $S_i$. If we find minimum energy consumption for nodes in different spheres, the lower bound on the energy consumption of a node in the whole network will be:

$$max\{m_1, ..., m_k\} \qquad (2)$$

The total number of packets received at each iteration at a node cannot exceed $N - s_0 - 1$. Likewise, the total number of packets transmitted cannot exceed $N - s_0$. So the upper bound on the energy consumption of a node in the whole network is:

$$r.(N - s_0 - 1) + t.(N - s_0) = (r + t).(N - s_0) - r \qquad (3)$$

Suppose each node has the exact same amount of energy $EE$. So the maximum number of iterations, $T_{max}$, a sensor network can perform within its lifetime is:

$$\frac{EE}{(r + t).(N - s_0) - r} \leq T_{max} \leq \frac{EE}{max\{m_1, ..., m_k\}} \qquad (4)$$

## 3 Modeling Per-packet Energy Consumption

The energy consumed by a sensor node while sending, receiving (and even discarding) a packet can be described using a linear equation proposed in [4]:

$$Energy = m \times size + b \qquad (5)$$

The idea of this per-packet modeling is that there is a fixed component associated with device state changes and channel acquisition overhead ($b$), and an incremental component which is proportional to the size of the packet ($m \times size$). Experimental results confirm the accuracy of this equation. For various applications we can find the specific coefficients $m$ and $b$.

This model does not consider energy consumed in unsuccessful attempts to acquire the channel, or in messages lost due to collision, bit error or loss of wireless connectivity.

In this work, we only focus on unicast traffic. While working in unicast mode, a sensor mote overhears all traffic sent by nearby motes. Thus it is important to consider the energy consumption when a node determines that it is not the intended destination of received unicast traffic and discards the received packet. Since such computation is application specific (depends on the distance between sensor motes and their radio sphere of influence, etc.), we ignore this case, but the load of this energy can easily be added to the computation to provide more precise results.

Having unicast transmission in mind, per-packet energy consumption to send and receive, based on the linear equation, will be as follows ($E(operation\ x)$ means the energy needed to do the operation $x$):

$$E(send) = m_{send} \times size + b_{send} \tag{6}$$

$$E(receive) = m_{receive} \times size + b_{receive} \tag{7}$$

Available sensor motes in the market usually feature radios which are IEEE 802.15.4 compliant. This standard defines the protocol and interconnection of devices via radio communication in a personal area network (PAN) called LR-WPAN (Low-Rate Wireless Personal Area Network). The standard uses carrier sense multiple access with a collision avoidance (CSMA/CA) medium access mechanism and supports star as well as peer-to-peer topologies.

The LR-WPAN defines four frame structures:
1. A beacon frame, used by a coordinator to transmit beacons
2. A data frame, used for all transfers of data
3. An acknowledgment frame, used for confirming successful frame reception
4. A MAC command frame, used for handling all MAC peer entity control transfers

Beacon frames and MAC command frames are used for management and control purposes. For simplicity we don't consider these two frames and focus our efforts on the other two frames.

Figure 2 [5] shows the structure of the data frame, which originates from the upper layers.

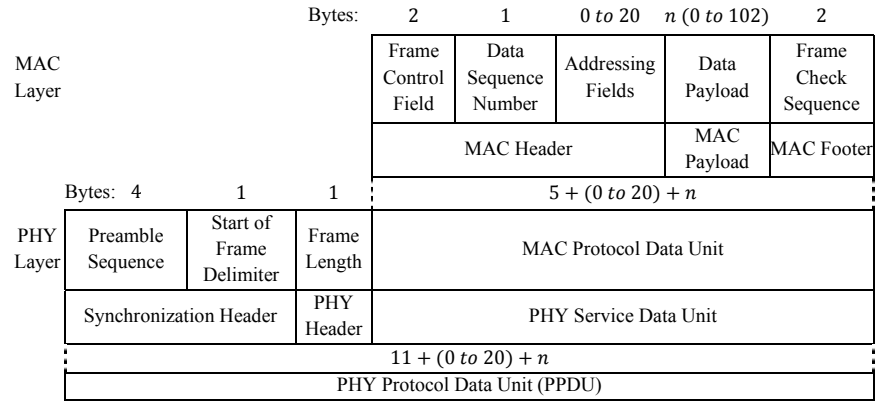

**Fig. 2.** Schematic view of the data frame [5]

The length of the physical data packet (PPDU) is $11 + (0\ to\ 20) + n$ bytes. The addressing fields shall comprise the destination address fields and/or the source address fields, dependent on the settings in the frame control field.

| 0/2 bytes | 0/2/8 bytes | 0/2 bytes | 0/2/8 bytes |
|---|---|---|---|
| Destination PAN identifier | Destination address | Source PAN identifier | Source address |

**Fig. 3.** Addressing fields

Practically 16-bit (2-byte) addresses i.e. short addresses are assigned to the sensor motes, source PAN identifier is left empty, and 2-byte destination PAN identifier is assigned. Altogether 6 bytes is needed for addressing purposes, hence the length of the data packet gets $17 + n$ bytes.

Figure 4 [5] shows the structure of the acknowledgment frame.

| 4 bytes | 1 byte | 1 byte | 2 bytes | 1 byte | 2 bytes |
|---|---|---|---|---|---|
| Preamble Sequence | Start of Frame Delimiter | Frame Length | Frame Control Field | Data Sequence Number | Frame Check Sequence |
| Synchronization Header | PHY Header | | MAC Header | | MAC Footer |

**Fig. 4.** Schematic view of the acknowledgment frame [5]

The length of the acknowledgment packet is 11 bytes. Although use of acknowledgment frames is optional, we will consider them in our computations for transfer reliability.

Altogether to transmit an $n$-byte packet, we need to send $17 + n$ bytes (data frame) and receive 11 bytes (acknowledgment frame). To receive this packet, we need to receive $17 + n$ bytes (data frame) and send 11 bytes (acknowledgment frame).

In our computations, we ignore beacons which are transmitted by the coordinator to provide synchronization services in IEEE 802.15.4. If beacons are not used in the PAN, the MAC sub-layer shall transmit using the unslotted version of the CSMA/CA algorithm. Briefly, the unslotted CSMA/CA algorithm has the following steps [5]:
1. Initialization of local variables related to backoff
2. Random backoff period
3. Clear Channel Assessment (CCA)
4. Transmission
5. Acknowledgment

The first two steps are internal operations and as noted before consume negligible amount of energy compared to communication operations. Thus, we do not include these two steps in the measurements. Also, an ideal network with complete symmetry is assumed without the need for extra prepended preambles or consecutive packet repetition in the sender's side or long listening period in the receiver's side.

In order to measure the lower and upper bounds of the energy consumption of routings based on the framework of IEEE 802.15.4 for a network in which every node transmits a packet with size $n$ at each iteration, we proceed as follows:

$$m_{send} = E(send\ 1\ byte) \tag{8}$$

$$b_{send} = E(CCA) + E(send\ 17\ bytes) + E(receive\ 11\ bytes) \tag{9}$$

$$m_{receive} = E(receive\ 1\ byte) \tag{10}$$

$$b_{receive} = E(listening) + E(receive\ 17\ bytes) + E(send\ 11\ bytes) \tag{11}$$

(An important note is that there is no listening before receiving the acknowledgment frame by the transmitter of the data frame, and no CCA before sending the acknowledgment frame by the receiver of the data frame.)

$$E(send) = m_{send} \times n + b_{send} \tag{12}$$

$$E(receive) = m_{receive} \times n + b_{receive} \tag{13}$$

$$m_i = \left(\frac{N - b_i}{s_i}\right) \times E(receive) + \left(\frac{N - b_i + s_i}{s_i}\right) \times E(send) \tag{14}$$

$$\frac{EE}{\left[\left(E(receive) + E(send)\right) \times (N - s_0)\right] - E(receive)} \leq T_{max} \leq \frac{EE}{max\{m_1, \dots, m_k\}} \quad (15)$$

## 4 Measurement of Bounds Using Tmote Sky Sensor Motes

Tmote Sky is an ultra low power wireless module for use in sensor networks, monitoring applications, and rapid application prototyping. Tmote Sky features the Chipcon CC2420 radio for wireless communications. The CC2420 is an IEEE 802.15.4 compliant radio providing the PHY and some MAC functions [6].

In this section, we will measure the lower and upper bounds of energy consumption for a network composed of Tmote Sky Sensor motes. To do so, we need some precise power measurements of different operations done by CC2420. We program our tests using TinyOS version 2.1.0 and nesC language. We repeat these measurements over 30 times and the average of the results is considered.

There is one byte difference between the packet size of TinyOS and the standard 802.15.4. TinyOS adds the field "TinyOS_IP" which is not defined in 802.15.4. Thus the length of the data frame is $18 + n$ bytes in TinyOS.

### 4.1 Measurement Setup

Our measurement setup is depicted in Figure 5.

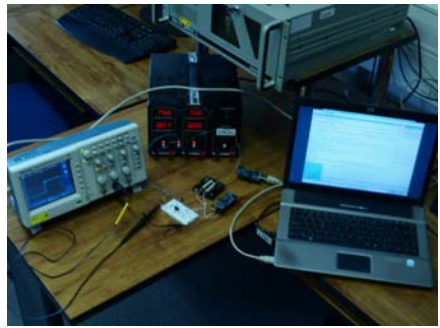

**Fig. 5.** Our hardware measurement configuration

The oscilloscope Tek TDS2012B [7] was chosen for our measurements which has a USB port in order to record precise data digitally. A 1.7 $\Omega$ test resistance was inserted in series between the power supply (3 $V$) and the mote. The two ends of the resistor were connected to the inputs of an instrumentation amplifier. An AD620AN instrumentation amplifier [8] with the gain of 98 was used to amplify the voltage across the resistor. To calculate the consumed power, we do as follows:

$$P = \frac{V_{scope}}{98 \times 1.7} \times 3 \times \Delta t_{scope} \quad Joule \tag{16}$$

### 4.2 CCA and Sending

The CC2420 has programmable output power. In our measurements, we used the maximum output power on CC2420 (0 dBm).

To measure the power needed for sending operation, a 46-byte packet (28 bytes data payload and 18 bytes header and footer) was sent. The scope trace is seen in Figure 6 (left).

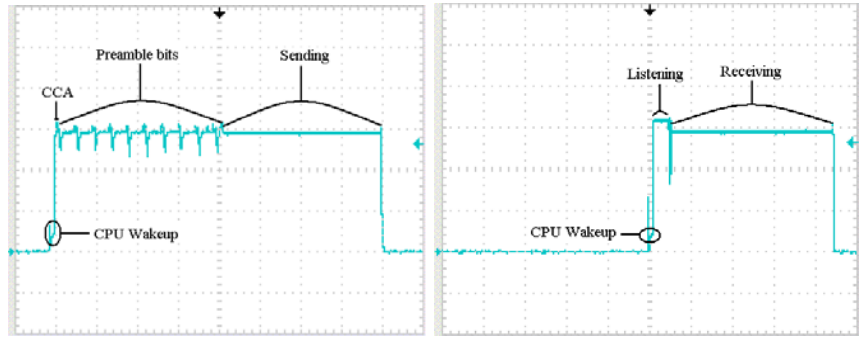

**Fig. 6.** Sending (left) and receiving (right) operations. X-axis: 25 mS, Y-axis: 1 V per division.

Voltage for CCA operation is $3.2\ V$ and time is $1.4\ mS$, hence $P = 0.08\ mJ$.

As noted before, we ignore extra preamble bits and only consider the 4-byte preamble sequence at IEEE 802.15.4 standard data frame. These 4 bytes are among the extra ones in the area marked "Preamble bits" in Figure 6 (left). It takes $96\ mS$ with the voltage $2.92\ V$ to send 42 bytes $(46 - 4)$. Thus, power consumptions to send 1 byte, 11 bytes and 18 bytes are $0.12\ mJ, 1.32\ mJ$ and $2.16\ mJ$, respectively.

### 4.3 Listening and Receiving

To measure receiving power consumption, a 46-byte packet is received. The scope trace is depicted in Figure 6 (right).

There is listening for the duration of $10\ mS$ which is a short periodic receive check before receiving data. Voltage for listening is $3.2\ V$, hence $P = 0.58\ mJ$.

The time for receiving operation is roughly the same as sending operation for transmission of a packet, but the voltage decreases to $2.88\ V$. Therefore, the power consumptions to receive 1 byte, 11 bytes and 18 bytes are $0.12\ mJ$, $1.3\ mJ$ and $2.13\ mJ$, respectively.

### 4.4 Results

Based on the measurements in previous section, we rewrite the equations 12 and 13:

$$E(send) = 0.12 \times n + 3.54 \quad mJ \tag{17}$$

$$E(receive) = 0.12 \times n + 4.03 \quad mJ \tag{18}$$

### 4.5 Example

Consider a network consisted of 29 nodes as depicted in Figure 1. The nodes are Tmote Sky sensor motes (with CC2420 radios), and in the intervals of 10 seconds send packets with 2-byte data payload. In each iteration, every node transfers its own data plus the data coming from the outer spheres (but leaves that only transfer their own data) to the inner spheres, and finally all the packets are collected at the base station. To calculate the lifetime bounds, we work as follows (Tmote Sky sensor motes use 2 AA batteries with the total amount of 30780 Joules):

$E(send) = 0.12 \times 2 + 3.54 = 3.78 \quad mJ$
$E(receive) = 0.12 \times 2 + 4.03 = 4.27 \quad mJ$
$s_1 = 4, b_1 = 5, m_1 = 52.08 \quad mJ$
$s_2 = 6, b_2 = 11, m_2 = 27.93 \quad mJ$
$s_3 = 10, b_3 = 21, m_3 = 10.22 \quad mJ$
$s_4 = 8, b_4 = 29, m_4 = 3.78 \quad mJ$
$max\{m_1, \ldots, m_4\} = 52.08 \quad mJ$

$$\frac{30780}{221.13 \times 10^{-3}} \leq T_{max} \leq \frac{30780}{52.08 \times 10^{-3}} \rightarrow 139194 \leq T_{max} \leq 591014$$

Considering the 10-second intervals, the lower lifetime bound will be about 387 hours and the upper lifetime bound 1642 hours.

In this example, the bottle neck is $S_1$, the sphere which in fact determines the upper bound. By better positioning the nodes, it is possible to spread the load of packet transmissions approximately equally on all the nodes and improve the upper bound.

In case the base station does not need hard real-time evaluation of the data sent by the nodes, it is possible for the nodes to aggregate their data and send it in longer intervals. This way we can improve the bounds significantly. Imagine in the previous example the nodes instead of sending packets consisting 2-byte data payload every *10 S*, aggregate data and send packets with 6-byte data payload every *30 S*. The lower bound will be 1036 hours and the upper bound 4398 hours. Comparing with the previous result, the bounds are now about 3 times more. The reason is obvious; the fixed component of the linear equation, which determines the power consumed to acquire the channel, and to send and receive acknowledgment frame and header and footer of the packet, is much higher than the incremental component for CC2420.

## 5   Conclusions

In this paper we presented how to compute the lower and upper lifetime bounds of WSNs. Although we concentrated on IEEE 802.15.4 frames, this method can be extended to any type of frame used in the other standards for wireless data transmission. We did not discuss the effect of different Network topologies supported in 802.15.4 (star or peer-to-peer) in our measurements. Also we did not take into consideration the presence of coordinators and beacons and superframe structures. Adding them will lower the lifetime bounds. However to provide more precise bounds, it is desirable to work on these concepts in future work.